\documentclass[12pt]{article}

\usepackage{amsmath,amsfonts,epsfig,color,latexsym,graphicx}

\newcommand{\be}{\begin{equation}}
\newcommand{\ee}{\end{equation}}
\newcommand{\bea}{\begin{eqnarray}}
\newcommand{\eea}{\end{eqnarray}}

\let\newsection=\section
\renewcommand{\section}{\setcounter{equation}{0}\newsection}

\begin{document}

\begin{flushright}
TIT/HEP-588
\end{flushright}
\vskip.5in

\begin{center}

{\LARGE\bf Diffractive Vector Meson Photoproduction from Dual 
String Theory }
\vskip .5in
\centerline{\Large Peter G. O. Freund$^1$ and Horatiu Nastase$^2$}
\vskip .5in

\end{center}
\centerline{\large $^1$ Enrico Fermi Institute and Department of Physics,} 
\centerline{\large University of Chicago, Chicago, IL 60637, USA}
\vspace{.5cm}

\centerline{\large $^2$Global Edge Institute, Tokyo Institute of Technology}
\centerline{\large Ookayama 2-12-1, Meguro, Tokyo 152-8550, Japan}

\vskip .4in

\begin{abstract}

{\large We study diffractive 
vector meson photoproduction using string theory via AdS/CFT.
The large $s$ behavior of the cross sections for the scattering of the vector meson $V$ on a proton 
is dominated by the soft Pomeron,
$\sigma _V\sim s^{2\epsilon-2\alpha'_P/B}$, 
where from the string theory model of \cite{nastase2}, $\epsilon$ is approximately
1/7 below 10 GeV, and 1/11 for higher, but still sub-Froissart, energies. This is due to the production of 
black holes in the dual gravity. 
In $\phi$-photoproduction the mesonic Regge 
poles do not contribute, so that we deal with a pure Pomeron contribution. This allows for an 
experimental test.
% which confirms the dual black hole picture.
At the gauge theory "Planck scale" of about 1-2 GeV, 
the ratios of the soft Pomeron 
contributions to the photoproduction cross-sections of different vector mesons involve not only 
the obvious quark model factors, but also the Boltzmann factors
$e^{-4 M_V/T_0}$, with $T_0$ the temperature of the 
dual black hole. The presence of these factors is confirmed in the 
experimental data for $\rho, \omega, \phi, J/\psi,$ and $\psi(2S)$ photoproduction and is compatible with the meager 
$\Upsilon$ photoproduction data. Throughout, we use vector meson dominance, and from the 
data we obtain $T_0$ of about $1.3~ GeV$, i.e. the gauge theory "Planck scale,"
as expected. The ratio of the experimental soft Pomeron onset scale $\hat{E}_R\sim 9$ GeV and of the 
gauge theory Planck scale, $T_0 \sim 1.3$ GeV conforms to the  
theoretical prediction of $N_c^2/N_c^{1/4}$.}

\end{abstract}

\newpage

\section{Introduction}

AdS/CFT \cite{malda} started as a duality between the conformal $SU(N)$ 
Super-Yang-Mills field theory in 4 dimensions at large $N$, and string theory 
in the curved 10-dimensional space, $AdS_5\times S_5$, in the low energy supergravity 
limit. The power of this approach derives from the fact that it relates a strong 
coupling theory --- $SU(N)$ SYM at large $N$ and large $g^2_{YM}N$ --- to a weakly coupled
string theory, where one can calculate. In \cite{bmn} the duality was 
extended to actual string theory, away from the supergravity limit, but 
only for  the gauge theory sector involving operators with very large R charge.

The duality was extended to various theories with less supersymmetry and/or
conformal invariance. Interesting examples are given e.g., by \cite{ks,mn}. 
But the difficulty in applying AdS/CFT to real QCD stems not only from the 
absence of a gravity dual to a nonsupersymmetric gauge theory with light 
quarks. The problem is that a calculable version with broken supersymmetry, with $N_f/N_c$ fixed, with light quarks  
has not yet been found, much as one can impose each of these  
requirements by itself.  Moreover in real QCD $N=3$ and $g^2_{YM}N$ are both finite.
In QCD, $g^2_{YM}$ runs, but $g^2_{YM}N$ never becomes large enough. 
In AdS/CFT the $1/N$ and $1/(g^2_{YM}N)$ corrections 
are mapped to string $\alpha '$ and $g_s$ corrections. String $\alpha '$
corrections are worldsheet corrections, i.e. 2-dimensional field theory quantum corrections,
whereas $g_s$ corrections are string spacetime quantum corrections. 

String 
theory is defined only perturbatively, with nonperturbative definitions 
available only in special cases. Moreover, string theory calculations 
in the strong coupling regime are hard. Thus, in order to apply AdS/CFT to real QCD, 
we need to first make sure that string corrections are small in the dual theory. That 
is in general not true, as finite $N$ and $g^2_{YM}N$ translate into large 
quantum corrections in string theory.

Recently, a lot of work has been devoted to gravity dual models of QCD for phenomenological purposes, most notably 
the Sakai-Sugimoto model \cite{ss}, but in that case we have effectively a quenched approximation, as the 
dual gravity background contains
no back-reaction of the D8-brane where the quarks live, thus effectively one works with $N_f/N_c
\rightarrow 0$. Another popular way of applying gravity dual results to describe RHIC physics \cite{RHIC}, 
uses ${\cal N}=4$
SYM models at finite temperature, corresponding to a static black hole in $AdS_5$. While
in these models one can perform explicit calculations, it is not completely 
clear why one should be allowed to use ${\cal N}=4$ SYM instead of non-supersymmetric
QCD, even if both are at finite temperature. Moreover, neither the string theory 
quantum corrections, nor the corresponding QCD $1/N$ and 
$1/(g^2_{YM}N)$ corrections, are under control.

However, in \cite{kntwo,knthree,nastase2,nastase3} it was noticed that there exists a particular
class of problems for which quantum corrections in the dual theory are small,
and one can use AdS/CFT for real QCD as well. The case in point is that of 
soft high energy scattering. It was shown that the total QCD cross section 
at large center-of-mass energy squared $s$ can be calculated, and the behaviour
$\sigma _{tot}\sim s^{\epsilon}$ and its later unitarization to $\ln^2(s/s_0)$ can be derived
and matched against experiment. In the gravity dual this corresponds to the
creation of black holes, which in the RHIC energy regime can be mapped to 
the fireball observed in the nucleus-nucleus collisions \cite{nastase3,nastase4}. 

A natural next step is to consider diffractive 
vector meson photoproduction, which is governed by the same soft Pomeron 
physics, and for which a wealth of  experimental data is available. 
In this paper, we take this step, and find that 
$\phi$-photoproduction which is known \cite{freund2} 
to provide the cleanest test of soft Pomeron behaviour,
matches well with the experimental evidence. We also analyze the ratios 
of the production cross sections of various vector mesons in the soft regime, and find 
a formula which reproduces the data for all known mesons, and allows us to
extract from experiment a ``gauge theory Planck scale'' of about 
$1.3$ GeV. 

In section 2 we describe the gravity dual picture for the soft physics.
In section 3 we present the general picture of diffractive vector meson 
photoproduction and show how to extract the soft Pomeron contribution. 
In section 4 we discuss the gravity dual picture for  diffractive vector meson 
photoproduction and compare it with experimental data. Finally, in section 5 we 
present our conclusions.

\section{Black hole production in the gravity dual and the Pomeron}

At energies above a gravitational theory's Planck scale, i.e. at $\sqrt s>M_{Pl}$, scattering
of any two particles should produce black holes. The hard, large $t$-scattering, however, 
will still be governed by string amplitudes, since a black hole radiates 
particles thermally, at generically low energies. On the other hand, soft small $t$-scattering, 
or processes with many particles and small average emitted energy in the final state,
will be dominated by black hole creation.

In \cite{kn} the process of black hole creation was analyzed using an old 
idea of 't Hooft \cite{thooft}, 
modelling the two colliding particles by Aichelburg-Sexl (AS)
gravitational shockwaves \cite{as}. 
\be
ds^2= 2dx^+dx^--\Phi(x^i)\delta (x^+)(dx^+)^2+d\vec{x}^2.
\ee
According to an argument of 't Hooft, quantum 
gravity corrections are negligible because massive interactions have finite 
range, but at small distances the gravitational shockwave gives a diverging 
time delay for the interaction ($\Phi(r\rightarrow 0)\rightarrow \infty$), 
making the corrections irrelevant, whereas for the massless interactions,
classical gravity reproduces the correct physics. It would be important to obtain a definitive 
proof of this statement. 
In \cite{kn}, the cross section for 
black hole formation in the  higher dimensional curved space scattering
was calculated, by scattering two AS waves,
based on earlier work \cite{eg} in flat 4-dimensional space. String corrections were 
also calculated, by scattering string-corrected AS shockwaves. It was 
found that in flat 4-dimensional space string corrections are exponentially
small at energies beyond $E_0\sim M_P^2/M_s$, which is of the order of the Planck scale
if the Planck and string scales are not too far apart.

On the other hand, AdS/CFT \cite{malda}
relates gauge theories living on systems of D-branes
with gravitational theories. In particular, a 4-dimensional conformal field theory 
will have a gravity dual of the type
\be
ds^2= \frac{\bar{r}^2}{R^2}d\vec{x}^2+\frac{R^2}{r^2}d\bar{r}^2+R^2 ds_X^2
=e^{-2y/R}d\vec{x}^2+dy^2+R^2 ds_X^2. 
\ee
Here the first two terms give the line element of the $AdS_5$ space of size $R$, and the last term gives 
the line element of the $S^5$ of radius R.

A nonconformal theory will have a gravity dual modified in the IR, i.e. small $\bar{r}$ or large y. 
In the simplest model of this type, one describes the unknown 
modification by an "IR brane,'' a cut-off at $\bar{r}_{min}\sim R^2\Lambda_{QCD}$
with 
$\Lambda_{QCD}$ the lightest excitation of the theory \cite{ps}. 

This simple model is dual to a pure glue gauge theory. To model the pion, the lightest excitation of real QCD, 
which is
a $q\bar{q}$ state, one assumes the position of the cut-off to be a dynamical brane. Its fluctuation, the radion, 
corresponds
to a simple scalar "pion" \cite{gid}.

High energy scattering in the gauge theory of modes with momentum $p$ and 
wavefunction $e^{i px}$ was mapped in \cite{ps} to scattering in 
the gravity dual with local AdS momentum $\tilde{p}_{\mu}=(R/\bar{r})p
_{\mu}$ and wavefunction $e^{ipx} \psi (r, \Omega)$. The gauge amplitude
is given by the gravitational amplitude, integrated over the extra coordinates, 
and convoluted with the wavefunctions. The string tension 
$\alpha ' = R^2/(g_sN)^{1/2}$ corresponds to the gauge theory string 
tension $\hat{\alpha '}= \Lambda _{QCD}^{-2}/ (g^2_{YM}N)^{1/2}$, and
\be 
\sqrt{\alpha '}\tilde{p}_{string}\leq \sqrt{\hat{\alpha}' }p_{QCD}.\label{stringcorresp}
\ee

This formalism was applied in \cite{kntwo,nastase2} to the case of soft 
scattering with black hole formation. It was found that  
most of the integral over the coordinate $\bar{r}$ is supported in the 
IR region of small $\bar{r}$, close to the cut-off. For self-consistency though,
it still has to be supported  away from $\bar{r}_{min}$. The cross section 
for black hole creation in the gravity dual was calculated in the classical
shockwave scattering picture mentioned before, and then using a simple eikonal 
model, an elastic $2 \rightarrow 2$ quantum amplitude was derived, and used in the 
Polchinski-Strassler formalism \cite{ps}. Let us emphasize again that 
in the soft scattering regime, the black holes are
produced on the average close to the IR cut-off i.e to the IR brane, but still away 
from it.

In the gravity dual, there are three energy scales of interest: the Planck 
scale $M_P$, the scale $E_R=N^2 R^{-1}$ at which the produced black holes reach the AdS size, 
and the scale $E_F$ at which the black holes are big enough to reach the IR brane. 
This last scale clearly cannot be calculated without knowing the details of the 
IR modification of the gravity dual to a given gauge theory. The simple 
cut-off model cannot be used to calculate it. According to Eq. (\ref{stringcorresp}), 
in gauge theory these energy scales correspond 
to $\hat{M}_P= \Lambda_{QCD} N^{1/4}$, $\hat{E}_R= \Lambda
_{QCD} N^2$, and an unknown $\hat{E}_F$. It was found \cite{kntwo,nastase2,gid} that between $\hat{M}_P
$ and $\hat{E}_R$, when in the gravity dual one produces black holes 
small enough to be considered approximately in flat space, the gauge 
theory cross section grows like
\be
\sigma_{gauge, tot}\simeq K (s/s_0)^{\epsilon}= K (s/\hat{M}_P^2)^{1/7}.
\ee
where $K$ is a constant, as are $\bar{K}$ and $K'$ in the next two equations. 
Between the energies $\hat{E}_R$ and $\hat{E}_F$, in the gravity dual the black holes
are large enough to feel the AdS size, but not the size of $X_5$. Therefore 
they grow in $AdS_5\times X_5$ with $X_5$ large, giving the gauge theory cross 
section \cite{nastase2}
\be
\sigma_{gauge, tot}\simeq \bar{K} (s/\bar{s}_0)^{\epsilon}= 
\bar{K} (s/\hat{E}_R^2)^{1/11}
\ee
Above $\hat{E}_F$, in the gravity dual the black holes become so large that they reach the IR brane. The 
gauge theory cross section then saturates the Froissart bound 
\cite{kntwo,knthree}, \cite{gid}, \cite{frois}, \cite{heis}.
\be
\sigma_{gauge, tot}\simeq K' \frac{\pi}{M_1^2} ln^2 \frac{s}{s_1}
\ee
where the mass $M_1$ corresponds to the lightest state in the theory. 
In the simple cut-off model, this is the 
mass gap, or the lightest glueball in the gauge theory. In QCD, where the pion is a pseudo-Nambu-Goldstone boson, 
$M_1$ is replaced by the pion mass
$m_{\pi}$. As mentioned, this Nambu-Goldstone boson is modeled by the radion, or equivalently, by the position 
of the IR brane in the gravity dual. There is then also a different
Froissart 
onset scale $E_F'\neq E_F$, and therefore $\hat{E}_F'\neq \hat{E}_F$ in the gauge theory. This corresponds to
the scale at which brane bending reaches the black hole.

As mentioned in the introduction, for this analysis to apply at finite $N$ and finite $g^2_{YM}N$, 
as  in real QCD, we have to make sure that string corrections are small.
They are not small for hard scattering, but for soft scattering at $t$ fixed and $s\rightarrow 
\infty$, these string corrections are small above an energy scale $\sim M_P^2/M_s$ for flat
4-dimensional space. If 't Hooft's argument is valid, this should be actually
around $M_P$. Making use of the optical theorem $\sigma _{total}(s)
={\rm Im}A(s,t=0)/s$, the same analysis holds for the total cross section.

Note that string corrections to the background itself will be large, and this 
will translate into modifications of the various energy scales, in particular of
$\Lambda_{QCD}$, the dual of the AdS size $R^{-1}$. A priori, this also entails modifications of the ratios
$\hat{M}_P/\Lambda_{QCD}, \hat{E}_R/\Lambda_{QCD}$, and of $\hat{E}_F$.

By contrast, string corrections to soft scattering in a given background {\em are} 
small, meaning that we can trust the cross section calculations.
This allowed for a successful match \cite{nastase2} of the ``soft Pomeron'' 
exponent $\sigma _{tot} \sim s^{1/11}\simeq s^{0.0909}$,
expected to set in at $N_c^2 M_{1,glueball}\sim 10 GeV$ with the energy dependence
$\sigma _{tot}\sim s^{0.0933\pm 0.0024}$, observed experimentally as of 9 GeV \cite{compas,pdgold}.

\section{Diffractive Vector Meson Photoproduction (DVMP)}

\subsection{Vector Meson Dominance Model of Vector Meson Photoproduction}

Before applying AdS/CFT ideas to diffractive vector meson photoproduction processes, let us briefly recall how the 
"soft Pomeron" dominates them. A particularly simple and intuitive picture \cite{freund1} for the 
photoproduction $\gamma p \rightarrow V p$ of the vector meson $V$ is obtained  
in the Vector Meson Dominance (VMD) approximation. One treats the photon $\gamma$ as undergoing a transition to 
the a virtual vector meson $V$ which then scatters elastically on the target proton. This involves the 
corresponding $\gamma \rightarrow V$ transition matrix element 
\be
<0|j_\mu(0)|V_a>=f_V m_V^2 \eta _{\mu a}        \label{f_V}
\ee
(with $\eta_{\mu a}$ the Minkowski metric), 
as well as a $V$-propagator evaluated on the photon's mass-shell. This propagator then cancels the factor 
$m_V^2$, used for notational simplicity in the definition of $f_V$ in Eq. (\ref{f_V}) for the transition matrix 
element. Also  for notational simplicity this definition already includes the electomagnetic coupling constant. 
This way the amplitude $A_{\gamma p\rightarrow Vp}(s,t)$, 
with $s$ and $t$ the usual Mandelstam variables, is given in terms of the $Vp$ elastic scattering amplitude
$A_{Vp}(s,t)$ as
\be
A_{\gamma p\rightarrow Vp}(s,t) = f_V A_{Vp}(s,t)
\ee

With VMD, understanding diffractive vector meson photoproduction reduces to understanding 
diffractive vector meson-proton elastic scattering and this calls for Regge theory, and 
specifically for Pomeron dominance.

\subsection{Regge theory for $Vp$ scattering}

Consider the amplitude ${A}_{Vp}(s,t)$ for $Vp\rightarrow Vp$ scattering. 
In gauge theories, it was found that one can generically describe the various Regge limits in terms of 
a tree level theory of effective particle ("Reggeon") interactions. 
In the case of QCD, the leading trajectory is the Pomeron. 
It is an effective particle made up of gluon interactions, and in the case of soft scattering, we have the 
"soft Pomeron." It was found that several effective particles can account for several regimes.
For $t\leq 0$ fixed and $s\rightarrow \infty$ Regge theory gives
\be
{A}_{Vp}(s,t)\sim \beta_V(t)sig_P(t)s^{\alpha_P(t)}\label{Vp}
\ee
where 
\be
\alpha_P(t)=\alpha_P(0)+\alpha'_Pt+...
\ee
is the Pomeron trajectory, and
\be
sig_P(t)= \frac{-1-e^{-i\pi \alpha_P(t)}}{\sin \pi \alpha_P(t)}
\ee
its signature factor.\footnote{The Pomeron has even signature. Other Regge trajectories can have odd signatures.}
The Pomeron Regge residue in $Vp$-scattering $\beta_ V(t)$ is real for $t\leq 0$. 
If $\alpha_P(0)\simeq 1$, as can be seen from  
total cross section data even at energies below the onset of the Froissart regime,  then $sig_P(t)\simeq i$.

The total cross section for $Vp$ scattering, $\sigma_{Vp,total}(s)$
is given by the optical theorem, 
\be
\sigma_{Vp,total}(s)=\frac{{\rm Im}{A}_{Vp}(s,t=0)}{s}\simeq \beta_V(0)s^{\alpha_P(0)-1}\equiv 
\beta_V(0)s^\epsilon.
\ee
Therefore $\epsilon=\alpha_P(0)-1$, and from the gravity dual description in section 2, we should have
$\epsilon\simeq 1/7$
below $10~GeV$, and $\simeq 1/11$ between $10$ GeV and $\hat{E}_F$, the onset of the Froissart saturation 
behavior.

On the other hand, the elastic differential cross section for $Vp\rightarrow Vp$ scattering is 
\be
\frac{d\sigma_{Vp}(s,t)}{dt}=\frac{1}{16\pi s^2} |{A}_{Vp}(s,t)|^2\simeq \frac{1}{16\pi}
|\beta_V(t) sig_P(t)|^2
s^{2(\alpha_P(t)-1)}\equiv F_V(t)s^{2\epsilon+2\alpha'_Pt+...}
\ee
For small $t$, such that $-1GeV^2<t\leq 0$, the prefactor is well approximated by an exponential in $t$,
\be
F_V(t)=\frac{1}{16\pi}|\beta_V(t)sig_P(t)|^2 \simeq F_V(0)e^{Bt}\label{expon}.
\ee
At large $|t|$ it becomes a power law.

By integrating the differential cross section over $t$ we obtain $\sigma_ {Vp,elastic}(s)$, 
the total elastic cross section.
Because of the the integrand's rapid exponential fall-off, one can extend the integral to $-\infty$. Thus
\be
\sigma_{Vp,elastic}(s)=\int_{-\infty}^0 dt \frac{d\sigma_{Vp}(s,t)}{dt}
\simeq\int_{-\infty}^0 dt F_V(t)s^{2\epsilon+2\alpha'_Pt+...}\simeq
\frac{F_V(0)}{B}s^{2\epsilon}(1+2\frac{\alpha'_P}{B}\ln s)^{-1}
\ee
On the other hand, $\alpha'_P/B< \epsilon\ll 1$, since roughly, $\alpha'_P\sim 0.2 GeV^{-2}$ and $B\sim 2-4 GeV^{-2}$), 
which means that we can bring the corresponding term into the exponent, and obtain
\be
\sigma_{Vp,elastic}(s)\simeq \frac{F_V(0)}{B}s^{2\epsilon-2\alpha'_P/B}\equiv \frac{F_V(0)}{B}s^{2(\alpha_P(<t>)-1)},
\ee
where by definition $<t> =-1/B$ is the average $t$ of the integral. 

%If the elastic cross section $\sigma_{Vp,elastic}(s)$
%would have dominated the total cross section $\sigma_{Vp,total}$, we would have deduced that 
%$2(\alpha_P(<t>)-1)=\alpha_P(0)-1=\epsilon$, i.e. that 
%\be
%-\alpha'_P/B=\alpha'_P<t>=-\epsilon/2,\label{exponent}
%\ee
%but since it doesn't, 
%we can't deduce this. Still, we expect $<t>$ to be close to $-\epsilon/(2\alpha'_P)$. In fact, we will see that 
%in the next subsection that such a value is experimentally plausible.

As we shall see, the experimentally observed value of $<t>$ is close to $-\epsilon/(2\alpha'_P)$. 
%which differs by a factor $\ln s/2$ from earlier estimates.

\subsection{Regge theory of DVMP}

We can now use vector meson dominance (VMD) to relate the analysis of the previous subsection to 
diffractive vector meson photoproduction (DVMP).

%Consider the amplitude for diffractive vector meson photoproduction, ${A}_{\gamma p
%\rightarrow Vp}(s,t)$. In this case, vector meson dominance refers to the statement that we can separate a transition
%element on the incoming leg, $\gamma \rightarrow V$, i.e. replace $\gamma $ with $V$ and multiply by 
%$f_V\equiv <0|j_{\mu}|V^a>$, with $j_{\mu}$ the electromagnetic current 
%and $|V^a>$ the vector meson state,
%\be
%{A}_{\gamma p\rightarrow V p}(s,t) \simeq <0|j_{\mu} |V^a> {\cal A}_{Vp}(s,t)\equiv f_V {\cal A}_{Vp}(s,t)
%\ee

Combining Eqs. (3.2) and (3.7), the differential photoproduction cross section is
\be
\frac{d\sigma _{\gamma p\rightarrow Vp}(s,t)}{dt}=\frac{1}{16\pi s^2} |{\cal A}_{\gamma p \rightarrow Vp}(s,t)|^2
=|f_V|^2\frac{d\sigma_{Vp}(s,t)}{dt}=|f_V|^2F_V(t)s^{2\epsilon+2\alpha'_Pt+...}
\ee
Integrating  over $t$ we get the total $V$ meson photoproduction cross section,
\be
\sigma_{\gamma p\rightarrow Vp}(s)\equiv \int_{-\infty}^0 dt \frac{d}{dt}\sigma_{\gamma p\rightarrow Vp}(s,t)
\simeq|f_V|^2\sigma_{Vp,elastic}(s)=|f_V|^2\frac{F_V(0)}{B}s^{2(\alpha_P(<t>)-1)}\label{sigmaV}
\ee

On the other hand, the Compton scattering amplitude ${A}_{\gamma p}(s,t)$ can be obtained by a double application of VMD. 
We can relate the elastic process $\gamma p\rightarrow \gamma p$ with the elastic process $Vp\rightarrow Vp$
by separating a transition element on both the incoming and the outgoing $\gamma$, and summing over all vector 
mesons $V$ , with the result
\be
{A}_{\gamma p}(s,t)\simeq \sum _V f_V {A}_{Vp}(s,t) f_V^*, %= \sum _V |f_V|^2{A}_{Vp}(s,t),
\ee
where we have discarded the much smaller off-diagonal $Vp\rightarrow V'p$ terms.
Making use at this point of the optical theorem, gives
\be
\sigma_{\gamma p, total}(s)=\frac{{\rm Im}{A}_{\gamma p}(s,t=0)}{s}\simeq
\sum_V|f_V|^2\frac{{\rm Im}{A}_{V p}(s,t=0)}{s}=\sum _V|f_V|^2\sigma_{Vp,total}(s),
\ee
so that
\be
\sigma_{\gamma p, total}(s)\simeq\left(\sum_V|f_V|^2\beta_V(0)\right)s^\epsilon\label{sigmatotal}
\ee

In (\ref{sigmaV}), the value of $<t>=-1/B$ is the same for all light $V$ mesons, 
but for each heavy meson both $<t>$ and $\epsilon=\alpha_P(0)-1$ take different values, and give a different exponent.
%Comparing (\ref{sigmaV}) with (\ref{sigmatotal}), if $\sigma_{\gamma p,total}$ could be approximated by 
%a sum over $\sigma_{\gamma p\rightarrow Vp}$ for light mesons $V$, then it would imply (\ref{exponent}). 
%However, experimentally the total cross section is much larger, so we cannot make the approximation. 
%But also experimentally, the exponent in (\ref{sigmaV}) for light mesons is very close to the exponent in 
%the total photoproduction cross section, (\ref{sigmatotal}), so we nevertheless have a good numerical
%approximation to $2(\alpha_P(<t>)-1)=\epsilon$, or (\ref{exponent}), for light mesons.

In order to emphasize the difference between the Pomeron parameters 
for heavy and for light mesons, we write $\alpha_{P,V}$ and $<t>_V$ in (\ref{sigmaV}), which thus becomes
\be
\sigma_{\gamma p\rightarrow Vp}(s)\equiv |f_V|^2\frac{F_V(0)}{B}s^{2(\alpha_{P,V}(<t>_V)-1)}\label{sigmaVV}.
\ee

Allowing for such a flavor-dependence in the Pomeron, 
heretic though it may seem from the point of view of Regge theory, has the virtue of agreeing with 
what is experimentally observed.  
Earlier work has introduced a, from the point of Regge theory both
acceptable and required, flavor-dependence of the Pomeron residues \cite{cf1, cf2}, 
but what we are doing here amounts 
to allowing a flavor dependence of what seems to be the very position  
of the Pomeron singularity in the complex angular momentum plane. Among other effects, this type of 
"Pomeron-flavoring" would destroy 
the factorization properties of the Pomeron.
This Pomeron-flavoring can be understood by taking note of the different kinematic regimes for 
light and heavy vector meson photoproduction. Because of the large masses of the $J/\psi, ~\psi(2S)$ and $\Upsilon$, 
large momentum transfers are set into play and one moves away from the diffractive soft Pomeron peak relevant 
for the photoproduction of the light vector mesons, into 
a region where the hard Pomeron takes over and where such a flavoring is not ruled out.

We now normalize the $s$ behaviour to a fixed $s_0$, and write
\be
\sigma_{\gamma p\rightarrow Vp}(s)\simeq |f_V|^2\frac{F_V(0)}{B}s^{2(\alpha_{P,V}(<t>_V)-1)}\equiv
|f_V|^2C_V (s/s_0)^{\tilde{\epsilon}_V}\label{scaling}
\ee
where $C_V$ depends on the choice of $s_0$ and the exponent of $s$ is called $\tilde{\epsilon}_V$. 
For the comparison of 
cross sections to have predictive power, the value $s_0$ must be singled out by the theory.
In the gravity dual calculation in the next section, $s_0$ will indeed be set at
$s_0\sim \hat{M}_P^2$, and we will predict ratios of $C_V$ for this value of $s_0$.

\section{Gravity dual picture and matching with experiment}

\subsection{Theory}

Let us now look at 
vector meson photoproduction from the dual point of view. As we saw, for all practical purposes, we are dealing 
with large $s$ $Vp$ scattering, a
soft Pomeron process. This means that it should be
related to black hole production. The total QCD cross-section is 
due to black hole production.

We are thus led to consider the process $p\bar{p}\rightarrow V\bar{V}$ around 
the gauge theory Planck scale.
As we argued in section 2, above the gauge theory Planck scale this process should be 
governed by black hole production in the gravity dual. Specifically,
a mini-black hole should be produced near the IR brane, i.e. at an 
average position $r_{av}$ away from the IR brane $r_{min}$, but close 
to it. Then this black hole
decays into particles. 

At this scale, one might wonder whether it is possible to even speak of a
black hole. Yet in string theory there are good reasons to believe that the 
production and decay of a black hole is a quantum process, and the apparent 
black hole loss of information due to the black hole temperature is just 
an illusion, i.e. the temperature is not indicative of information loss.
How this can happen in the case of a near extremal black hole has been discussed in \cite{malda2}.
Furthermore in \cite{nastase5} for a simple scalar field theory
toy model for the pion, a solution was found which mimics the properties of the gravity dual black hole, 
with an effective temperature and with {\em apparent}
information loss, even though the quantum field theory from which one started is unitary. 
Thus, although a pure quantum process takes place, the effect of the black hole being 
produced is simply to give an effective temperature. 

Therefore since the energy is small enough for the decay to produce 
on the average just two particles --- the least number needed for momentum conservation ---
the simplest decay mode of the Planck-sized black hole is into a 
particle-antiparticle pair. The decay of the dual black hole is 
mapped in QCD to a gluonic interaction followed by decay into vector mesons.
Thus the dual black hole's only observable effect in the QCD $p\bar{p}$ scattering, 
is the existence of well-defined relative probabilities of decay for various vector meson 
pairs, so that
\be
\frac{\sigma_{V\bar{V}}(t)}{\sigma_{V'\bar{V'}}(t)}\sim \frac{P_V}{P_{V'}}= e^{-\frac{2(M_V-M_{V'})}{T_0}}
\ee
where $t$ is the Mandelstam variable for the $p\bar{p}\rightarrow V \bar{V}$ "$t$-channel," 
$M_V$ and $M_{V'}$ are the vector meson masses, $T_0$ is the temperature of the 
Planck-sized ($M_P$) black hole,
and the factor $2$ in the exponent corresponds to the  
meson pair.  

In QCD this involves a highly nonperturbative process, so that the effect of the dual black hole would in 
principle be reproduced by very complicated QCD interactions. 

QCD being a quantum theory, the process is unitary, as we said, and
therefore the effect on the quantum amplitude for $p \bar{p}\rightarrow V \bar{V}$ 
is that it factorizes into a universal piece $a(s,t)$ independent of $V$, and a $V$-dependent Boltzmann factor,
\be
{A}_{p\bar{p}\rightarrow V\bar{V}}(s,t)\sim e^{-\frac{2M_V}{T_0}} a (s,t)
\ee
Let us now look at this process in the crossed $Vp \rightarrow Vp$ $s$-channel. If this 
is in the soft regime, due to gluon interactions, or equivalently 
``Pomeron exchange'', then the above picture should apply, and we have 
\be
{A}_{Vp}(s,t)\equiv {A}_{Vp\rightarrow Vp}(s,t)\sim e^{-\frac{2M_V}{T_0}} a(s,t)\label{thermal}
\ee
where $a(s,t)$ is a universal $2\rightarrow 2$ amplitude in the Regge regime. We can now use this relation in 
(\ref{Vp}) and (\ref{expon}) to obtain $\beta_V(t)=e^{-\frac{2M_V}{T_0}}\beta(t)$ and $F_V(t)=
e^{-\frac{4M_V}{T_0}}F(t)$.

Then, since the
process we are interested in is actually $\gamma p\rightarrow V p$, we can use VMD as in sections 3.1 and 3.2, 
obtaining the amplitude
\be
{\cal A}_{\gamma p\rightarrow V p}(s,t) \sim f_V e^{-\frac{
2M_V}{T_0}}a(s,t)
\ee
and the V meson photoproduction cross section 
\bea
\sigma_{\gamma p\rightarrow V p}(s)&\simeq& |f_V|^2  e^{-\frac{4M_V}{T_0}}C(s/s_0)^{2(\alpha_{P,V}(<t>_V)-1)}\nonumber\\
&=&|f_V|^2 e^{-\frac{4M_V}{T_0}}C(s/s_0)^{2\epsilon_V-2\frac{\alpha'_P}{B_V}}
\label{ratio}
\eea
or $C_V=Ce^{-\frac{4M_V}{T_0}}$ in (\ref{scaling}). This is our main result, which we will confront with 
experiment.

First however, let us go back and reexamine the assumption of the black hole whose 
temperature $T_0$ is of Planck size, and also understand the scale $s_0$ at which the comparison of ratios is to be 
made.

Why are Planck sized black holes with $T_0\sim M_P$ relevant? After all, we argued that for the
total QCD cross section in the gravity dual one has black holes of growing 
size, thus decreasing temperature, that stops at $T_{min}= a 4M_1/\pi$, (with $a$ a suitable numerical factor)
corresponding in QCD to $T= a 4m_{\pi}/\pi$ \cite{nastase3}. 
The growing size of the produced black holes is responsible for the 
$s^{\epsilon}$ behaviour of $\sigma _{tot}(s)$. 

But for the temperature
factor we have gone from the $p\bar{p}\rightarrow V\bar{V}$ amplitude to the 
$s$-channel, exchanging the original $t$ with $s$. The original $t$ had to be 
$\geq M_P$ to create black holes, but for $pV\rightarrow pV$ we need 
$|t|\leq M_P$, thus $t\simeq M_P$ for $p\bar{p}\rightarrow V\bar{V}$ and so indeed $T_0\sim M_P$.
A similar argument can be made exchanging $s$ with $t$, yielding a Planck scale value $s_0\sim \hat{M}_P^2$ of the 
Mandelstam variable $s$
at which the soft Pomeron behaviour sets in for the $pV\rightarrow p V$ $s$-channel.

Thus for V-photoproduction at $\sqrt{s}=\sqrt{s_0}\sim \hat{M}_P$, the soft Pomeron contribution, 
dual to black hole production,  should be 
distributed according to 
%(\ref{ratio}), 
temperature $T_0\sim \hat{M}_P$. Beyond that energy
one should have the 
$s^{\tilde{\epsilon}_V}$ behavior. For light vector mesons, with mass $M_V <\hat{M}_P$,
$\tilde{\epsilon}_V$ should be the soft Pomeron exponent $2\epsilon-2\alpha'_P/B$. As we saw, this is numerically close to
$\epsilon$.
On the other hand, for heavy vector mesons,
with $M_V>\hat{M}_P$, $\tilde{\epsilon}_V$ should be an exponent defined by hard 
scattering. The reason for that is purely kinematic, as it is well
known: the photon has $Q^2=0$, whereas the vector meson has 
$P^2=M_V^2$, as if $M_V^2$ had been effectively added to it. Thus, if $M_V>\hat{M}_P$,
the physics is no longer dominated by the lightest glueball. 

Also, for kinematic reasons, production of a heavy meson V  with $M_V>\hat{M}_P$ will start above
$\sqrt{s}_V = M_V$ instead of at $\sqrt{s_0}\sim \hat{M}_P$, and as we said will be dominated by hard scattering.
We expect that if we extrapolate the hard scattering formula $\sigma_{V,hard}
\sim s^{\tilde{\epsilon}_{hard}}$ down to $\sqrt{s_0}=\hat{M}_P$, the soft formula 
%(\ref{ratio}) 
should apply. In other words, the formula 
%(\ref{ratio}) 
has $\sqrt{s_0}\sim 
\hat{M}_P$ for heavy mesons also, even though it only applies for $s>M_V^2$.

\begin{figure}[bthp]

\begin{center}

\includegraphics[width = 0.6\textwidth]{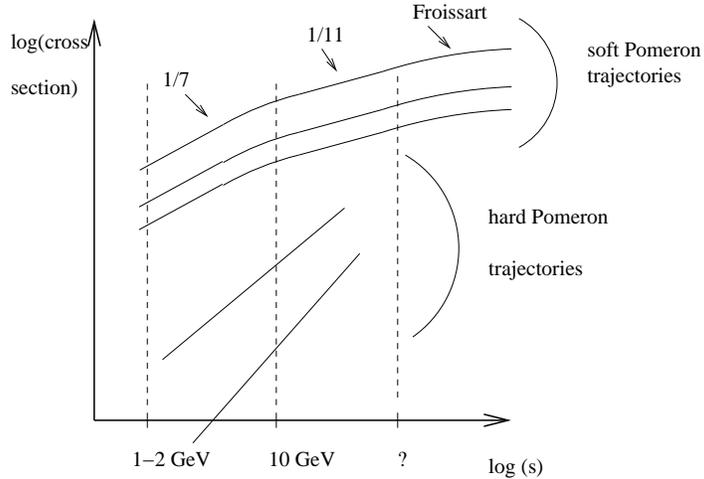}
\end{center}
\caption{
Predicted soft Pomeron contribution to the scattering cross 
section (due to production of dual black holes). The first energy scale 
is the gauge theory Planck scale, $\hat{M}_P=N_c^{1/4}M_{1,glueball}
\sim 1-2 GeV$. The second is $\hat{E}_R= N_c^2 M_{1, glueball}\sim 10 GeV$,
and the third is the \`{a} priori unknown Froissart scale, $\hat{E}_F$, above 
which $\sigma\sim log^2(s)$, thus $log(\sigma )\sim 2 log (log (s))$. 
For vector meson photoproduction, the exponent is $\tilde{\epsilon}=2\epsilon-2\alpha'_P/B$, 
which as was already mentioned, is seen to be close to the numerical value of $\epsilon$.
(the equality is true numerically, though not clear if true theoretically).
For vector mesons with masses
above $M_{1,glueball}$, we have hard scattering, with a larger exponent.
One could have a priori also a scale at which the exponent changes, as for 
the soft Pomeron exponent. At $\hat{M}_P$, the various mesons should be 
distributed according to the corresponding Boltzmann factors.}
\label{pomeron}
\end{figure}

In conclusion, the theoretical expectation for the Pomeron contribution to 
$\sigma_V$ is as sketched in fig.\ref{pomeron}. For light mesons, 
$M_V\leq \hat{M}_P$ we have the soft Pomeron exponent $\tilde{\epsilon}\equiv 2\epsilon-2\alpha'_P/B$, 
which as we saw is numerically close to $\epsilon$.
For heavy mesons $M_V
\geq \hat{M}_P$ we have hard the Pomeron exponent, but both the soft and hard 
Pomeron lines extrapolated down to about $\hat{M}_P$ should be distributed 
according to (\ref{ratio}).

\subsection{Comparison with Experiment}

Let us now compare the main formula 
(\ref{ratio}) 
to the experimental data, and extract the experimental 
value of $T_0$. For all light mesons, the fits show the  same soft Pomeron exponent, 
so that we can compare the ratios of cross sections %(\ref{ratio}) 
at any $s$, in particular large values of $s$, where neglecting Regge exchanges beyond the Pomeron is a good 
approximation.
By contrast,  different heavy mesons yield different Pomeron "trajectories," so that the powers of $s$ differ
and these different exponents are not predicted by theory. We will therefore extrapolate the data to the energy 
$\sqrt{s_0}\sim \hat{M}_P\sim 1-2GeV$, even though, as we already said, for heavy mesons the formula itself
starts applying only at a higher energy. We will then work out the relevant cross-section 
ratios from the so extrapolated data.

%We could in principle apply %(\ref{ratio}) 
%to the ratios of cross sections at some fixed high energy. Since for heavy 
%mesons the error bars on individual data points are very large however, that would be statistically unsound. The 
%clearly better procedure is to fit lines through the high energy data points for each vector meson (errors on the 
%line fits are smaller than individual errors)
%and extrapolate them down to $\sqrt{s}=\sqrt{s_0}\sim 1-2GeV$ (), and compare ratios there. 

Our procedure will amount to a simple graphical analysis taking advantage of best fit lines to the high energy data. 
A more extensive data analysis is not warranted at this point, since there are large
theoretical uncertainties. For instance, what is the precise value of $s_0$ at which we make the comparison, and 
what kind of corrections do we expect to the black hole exponential factor $e^{-4M_V/T_0}$? Given these uncertainties,
it makes little sense to try a more sophisticated statistical analysis of the data at this point.

The graph that we will be using for our analysis is Fig.2 of reference 
\cite{voss} (see also for instance, the review
\cite{kreisel}). The values of the exponents for the best fit lines at high energies are:
$\tilde{\epsilon}_{\rho}=\tilde{\epsilon}_{\omega}=\tilde{\epsilon}_{\phi}= 0.11$, $\tilde{\epsilon}_{J/\Psi}=0.41$, 
$\tilde{\epsilon}_{\psi(2S)}=0.55$ and $\tilde{\epsilon}_\Upsilon=0.9$.

Among the light mesons, 
$\phi$-photoproduction is the cleanest test of the soft Pomeron 
exponent, because in this case all mesonic Regge poles decouple \cite{freund2}. 
We therefore identify the cross-section curve for $\phi$-photoproduction with the soft Pomeron 
contribution. For the other light vector mesons, we look at data at higher energies,
$\sqrt{s}\gg 10 GeV$, where the Pomeron contribution dominates. We 
then extrapolate down to $\hat{M}_P$,
parallel with the $\phi$ line. The next cleanest line, also 
well split from $\phi$, is the $J/\psi$ line, where also all mesonic Regge poles decouple.
We therefore use the $J/\psi:\phi$ split to determine $T_0$ and then with the so obtained value of $T_0$ 
check the formula for the vector mesons for which the photoproduction cross section measurements carry larger errors. 
We also extrapolate the heavy meson lines 
down to $s_0\sim\hat{M}_P^2\sim 1-2~ GeV^2$. For concreteness, we at first choose $s_0=2 GeV^2$.

Numerically, as mentioned already, the energy dependence of total cross section is close to 
that of of the light meson photoproduction cross-sections. This way for light mesons 
$\tilde{\epsilon}\simeq \epsilon\simeq 0.11$,
even though, all one had a right to expect was $\tilde{\epsilon}=2\epsilon -2\alpha'_P/B$.  
In some sense it is as if we were testing the 
behaviour of $\sigma _V\sim s^\epsilon$. The fact that 
in the photoproduction cross-sections the switch from one power law to the other occurs around 
$9~GeV$ is particularly significant in that it does not depend on whether
$\tilde{\epsilon}\simeq \epsilon$ is true or not.

The $\phi$-photoproduction cross-section data 
have the general features that we expect. Most of 
the data are around 10 GeV, which from previous work \cite{nastase2} 
corresponds to the scale $\hat{E}_R$ at which in the dual gravity picture the black hole size reaches the AdS size.
This is the transition region between the $\epsilon=\frac{1}{7}$ and $\epsilon=\frac{1}{11}$.
Consequently, one measures mostly the energy dependence at 10 GeV, which should be of the form $s^{\epsilon}$ with
$\epsilon$ between $\frac{1}{7}$ and $\frac{1}{11}$, and indeed one 
finds $\epsilon=0.11\simeq (\frac{1}{7}+\frac{1}{11})/2$. Moreover, from the graph one can clearly see that $\epsilon$ 
is larger below 10 GeV. In fact one nicely fits the data with $\epsilon 
=0.15$, which is close to $\frac{1}{7}\simeq 0.143$.

Encouraged by this fit, let us 
turn to the formula 
%(\ref{ratio}) 
at $s=s_0$. The masses of the relevant mesons are
\bea
&& M_{\rho^0}= 775.8\pm 0.5 MeV;\;\; M_{\omega}= 782.59 \pm 0.11 MeV;\;\;
M_{\phi}= 1019.456 \pm 0.020 MeV\nonumber\\
&& M_{J/\Psi}= 3096.916\pm 0.011 MeV;\;\;
M_{\psi(2S)}= 3686.093\pm 0.034 MeV ;\nonumber\\&&
M_{\Upsilon}= 9460.30 \pm 0.26 MeV
\eea

For mesons that are close-by in mass and wave-functions, the relevant VMD factor 
$f_V^2$ should be dominated by the group 
theory factor $H_V \equiv (TrQ{\cal V})^2$, where in $(u,d,s,c,b)$ space,
 $Q= diag(2/3, -1/3, -1/3, 2/3, -1/3)$ the electric charge matrix
of the quarks, and ${\cal V}$ the quark content matrix of the vector meson. 
Since we have approximately 
\bea
&& \rho^0\simeq \frac{u\bar{u}-d\bar{d}}{\sqrt{2}},\;\;
\omega \simeq \frac{u\bar{u}+d\bar{d}}{\sqrt{2}}, \;\; \phi \simeq s\bar{s}
\nonumber\\&&
J/\psi \simeq c \bar{c}, \;\; \psi(2S)\simeq c\bar{c};\;\;
\Upsilon \simeq b\bar{b}
\eea

It follows that 
\bea
&& {\cal V}_{\rho^0}\simeq \frac{1}{\sqrt{2}}\tau_{3, (u,d)}, \;\;
{\cal V}_{\omega}= \frac{1}{\sqrt{2}}1_{(u,d)},\nonumber\\
&& {\cal V}_{\phi}=1_s, \;\; {\cal V}_{J/\psi}={\cal V}_{\psi(2S)} = 1_c ,\;\;
{\cal V}_{\Upsilon}=1_b
\eea
and so 
\be
H_{\rho^0}:H_{\omega}:H_{\phi}:H_{J/\psi}:H_{\psi(2S)}:H_{\Upsilon}= 9:1:2:8:8:2
\ee
The $\rho$ and $\omega$ mesons being almost mass-degenerate, the ratio of their photoproduction cross-sections 
at high $s$ 
should be 9:1, as has been known for a long time, see e.g. \cite{freund1} and can also be read 
off the experimental data.

%The $\omega:\phi$ split is within experimental errors at high energies
%-the isolated points at DESY energies on fig.2 of \cite{voss}.
%(even though from the graph we find at $\sim 10 GeV$ it is about 1/4 
%decade, i.e. a factor of 2 difference), so we will look at the 
%$\phi:J/\psi$ split to determine the temperature $T_0$ from the graph.

{\bf The $\phi:J/\Psi$ split.} If we extrapolate the $J/\psi$ photoproduction cross-section 
down to about 2 GeV, we find $\frac{\sigma_{\phi}}{\sigma_{J/\psi}}\sim 100$, which when compared with 
Eqs. (4.5) and (4.9) sets the temperature $T_0$ at $1.3$ GeV.
%\be
%\frac{\sigma_{\phi}}{\sigma_{J/\psi}}=\frac{2}{8} e^{\frac{4(M_{J/\psi}
%-M_{\phi})}{T_0}}
%\ee
%By comparing the the prediction to the observation, we then find $T_0\sim 1.3 GeV$.

{\bf The $\omega:\phi$ split}. With the temperature determined this way, we predict
\be
\frac{\sigma_{\omega}}{\sigma _{\phi}}=\frac{1}{2}e^{\frac{4(M_{\phi}-M_{\omega})}{T_0}}\sim 1.03
\ee
in agreement with experiment.

Here it is worth mentioning that the meson-dominated Pomeron picture \cite{cf1, cf2}, calls for a suppression of heavier
vector-meson photoproduction cross sections inverse proportional to the fourth power of the vector-meson mass. This 
replaces our Boltzmann factors. For the $\phi$ versus $\rho$ photoproduction, this means that the Boltzamnn 
factor $e^{\frac{4(M_{\phi}-M_{\rho})}{T_0}}\sim 2$ is to be replaced by
$\frac{M_\phi^4}{M_\rho^4}\sim 3$. Including the group theory factors, this yields 
$\frac{\sigma _{\rho}}{\sigma _{\phi}}\sim 13.5$, as opposed to our result of $9$. At high energies, 
where the soft Pomeron dominates, both pictures are compatible with
experiment, within the stated errors. 

{\bf The $J/\psi:\psi(2S)$ split.} Here $J/\psi$ and $\psi(2S)$ are 
 $n=1$, and $n=2$ $c\bar{c}$ $S-$states, 
respectively. Yet the corresponding $f_V M_V^2$ are not the same, as the 
$J/\psi$ and $\psi(2S)$ wavefunctions differ. 
The $n=2$ state will localize the quarks further from the origin, for 
instance. In order to extract the matrix element ratio, we use VMD, and start from the decay rates of 
$J/\psi:\psi(2S)$ into a lepton pair. In fact
\be
\frac{\Gamma_{J/\psi}(e^+e^-)}{\Gamma_{\psi(2S)}(e^+e^-)}
= \frac{|f_{J/\psi}|^2}{|f_{\psi(2S)}|^2}\frac{M_{J/\psi}}{M_{\psi(2S)}}
\ee
where the ratio of masses on the right hand side comes from the phase space factors of these
$S$-wave decays. Experimentally this ratio of widths is $5.37$ keV$/2.10$ keV $=2.55$. This way
\be
\frac{\sigma_{J/\psi}}{\sigma_{\psi(2S)}}|_{\sqrt{s}=\sqrt{s_0}\sim \hat{M}_P\sim 2 GeV}
=\frac{|f_{J/\psi}|^2}{|f_{\psi(2S)}|^2}
e^{\frac{4(M_{\psi(2S)}-M_{J/\psi})}{T_0}}
=10^{1.27} 
\ee

The experimental graph gives a split of about $10^{1.25}$.

{\bf The $J/\psi:\Upsilon$ split}. In this case the group theory and Boltzmann factors give 
\be
\frac{\sigma_{J/\psi}}{\sigma_{\Upsilon}}|_{\sqrt{s}=\sqrt{s_0}\sim 2 GeV}=\frac{8}{2}
e^{\frac{4(M_{\Upsilon}-M_{J/\psi})}{T_0}}\simeq 10^9
\ee
i.e. 9 decades difference. The experimental graph, blindly extrapolated to $2 ~GeV$ 
yields a difference of about 4.5 decades. 
But the $\Upsilon$ ``line'' consists of just 
2 close-by points with huge error bars, so that its slope could easily reach double the value suggested by
the best fit. The meson-dominated {\em soft} Pomeron picture is hardly applicable here, 
given the large momentum transfers kinematically required on account of 
the large $\Upsilon$ mass. This calls for the hard Pomeron in $\Upsilon$ photoproduction and eliminates 
the fourth-power mass dependence characteristic of the meson-dominated {\em soft} Pomeron picture, which
though valid for $\rho$ photoproduction, fails for 
$\Upsilon$ photoproduction.

A better determination of the $\Upsilon$ line through more accurate $\Upsilon$ photoproduction data 
will allow for an important additional test of these  ideas. 

On the whole, so far,  
our main formula with its characteristic Boltzmann factors compares quite well with experimental results.

%(\ref{ratio}) 

We were led to a Planck temperature of about 1.3 GeV by the data. This should be thought of as the 
QCD Planck scale $\hat{M}_P= N^{1/4}\Lambda_{QCD}$.

Also experimentally, the $\sigma_{tot}\sim s^{1/11}$ behaviour sets in
at about 9 GeV \cite{compas,pdgold}, which we can thus identify with $\hat{E}_R= N^2 
\Lambda_{QCD}$. Again, from the data we found  $\hat{E}_R/\hat{M}_P=
9 GeV/1.3 GeV$, which agrees embarrassingly well with the theoretical prediction $N^2/N^{1/4}= 9/1.316$ for $N=3$
colors. This is all the more mysterious, since, 
as mentioned in section 2, the values 
of energy scales can get possibly large string corrections. Could it be that in 
the ratios of energy scales, the string corrections cancel? 

We could have made a more rigorous analysis of the data, but as we already pointed out, 
there are several theoretical uncertainties, that could slightly 
modify the results, so it is not clear that it is useful getting better 
fits from the current experimental data, not until there is a better control on the theoretical 
uncertainties, even just in evaluating their size. 

On the theoretical side we have already mentioned the 
uncertainties in the exact value of $s_0$ and the unknown potential corrections to the
black hole Boltzmann factor. 
Also the hard Pomeron exponents $\tilde{\epsilon}_V$ for heavy vector mesons 
could change at a lower scale along our extrapolation. This could parallel what happens for 
the soft Pomeron in the gravity dual model. This would slightly affect
the extrapolation to $\hat{M}_P$. 

Then, the behaviour of the extrapolation of the hard 
Pomeron from $M_V$ down to $\hat{M}_P$ could also be slightly changed, 
as one needs at least $\sqrt{s}=M_V$ to even produce the mesons. For $J/\psi$ 
and $\psi(2S)$ that difference between the $\hat{M}_P$ and $M_V$ extrapolation 
is relatively small, but for $\Upsilon$ it could be larger, and it could contribute along with the 
large experimental errors to the observed discrepancy. Finally, we have 
assumed that the hard Pomeron  extrapolated down to $\hat{M}_P$ 
gives the soft Pomeron contribution at $\hat{M}_P$, but the soft Pomeron 
contribution could be slightly smaller.

\section{Conclusions}

In this paper we have analyzed diffractive vector meson photoproduction 
using AdS/CFT. The soft Pomeron behaviour of QCD was argued to be 
due to production of dual black holes. The photoproduction of $\phi$ meson
is the cleanest test of the soft Pomeron. Indeed the $\phi$ mass is sufficiently small for the applicability of
the soft Pomeron picture, and moreover mesonic Regge poles do not 
contribute to $\phi$ photoproduction \cite{freund2}. 

For vector meson photoproduction, the power $\tilde{\epsilon}=2\epsilon-2\alpha'_P/B$ differs
in principle form the power $\epsilon$ determined from the energy behavior of total cross sections, 
but they are numerically 
surprisingly close. 
Assuming outright equality, $\tilde{\epsilon}=\epsilon$, the $\phi$ data match  
the $s^{\epsilon}$-like theoretical prediction: $\sigma_V\sim s^{1/7}$ below 
$9$ GeV and $\sigma_V\sim s^{1/11}$ above $9$ GeV. The measured exponent of 0.11
is mostly due to data in the transition region around 9 GeV, and can be fit with an exponent of  
(1/7+1/11)/2=0.117. We emphasize though 
the absence of any theoretical understanding whatsoever for such an equality $\tilde{\epsilon}=\epsilon$ 
in the case of light vector mesons.

Then we have tested the general formula 
%(\ref{ratio}) 
at $s=s_0$.
The predicted ratios of soft Pomeron contribution  to $\sigma_V$ at about 1-2 GeV, 
were successfully compared  with experiment. This allowed us to extract the temperature $T_0\sim 1.3 GeV$. This 
$T_0$ has to be close to $\hat{M}_P$, and we therefore identified their values.  This led us to the value
$9 GeV/ 1.3 GeV$ for the ratio of the soft Pomeron onset scale, $\hat{E}_R$, to $\hat{M}_P$. As was pointed out, 
this agrees 
well beyond all expectation 
with the prediction $\hat{E}_R/\hat{M}_P=N_c^2/N_c^{1/4}=9/1.316$, though it is unclear why we have such 
good agreement, when a priori there could be dual string corrections.

An important further test of the ideas of this paper could be obtained from a more precise determination of the 
$s$ dependence of the $\Upsilon$ photoproduction cross section, allowing for a better determination of the 
extrapolated $J/\Psi:\Upsilon$ split at $s_0\sim 1-2 ~GeV^2$.

{\bf Acknowledgements}. We would also like to thank R. Enberg, 
whose graph we have used for a preliminary analysis, and
for pointing us to the correct sources for the data.
One of us (H.N.) would like to acknowledge valuable discussions with Chung-I Tan and the late Kyungsik Kang 
and thank the University of Chicago for the opportunity 
to visit which resulted in the beginning of this work. 
The research of H.N. has been supported in part by MEXT's program
"Promotion of Environmental Improvement for Independence of Young Researchers"
under the Special Coordination Funds for Promoting Science and Technology, and also with partial support from MEXT KAKENHI grant nr. 20740128.

\end{document}